\documentclass{JHEP3}
\usepackage{graphicx,amsmath,amssymb}
\usepackage{epsfig,multicol}
\usepackage{epsf}
\usepackage{epstopdf}
\input{epsf.sty}

\usepackage{graphicx}
\usepackage{amssymb}
\usepackage{amsmath}

\def\p{\partial}

\def\half{{1\over 2}}
\def\ka{\frac{k^2}{a^2}}
\def\({\left(}
\def\){\right)}
\def\[{\left[}
\def\]{\right]}

\def\e{\begin{equation}}
\def\q{\end{equation}}
\def\m{\begin{eqnarray}}
\def\n{\end{eqnarray}}

\title{Growth factor in f(T) gravity}
\author{Rui Zheng$^1$ \footnote{zhr315@mail.ustc.edu.cn} \ and Qing-Guo Huang$^2$ \footnote{huangqg@itp.ac.cn}
\\\small{\em
$^1$ Interdisciplinary Center for Theoretical Study, University of
Science and Technology of China, Hefei, Anhui 230026, China}
\\\small{\em
$^2$ Key Laboratory of Frontiers in Theoretical Physics, Institute of
Theoretical Physics, Chinese Academy of Sciences, Beijing 100190,
China} }

\abstract { We derive the evolution equation of growth factor for
the matter over-dense perturbation in $f(T)$ gravity. For instance,
we investigate its behavior in power law model at small redshift and
compare it to the prediction of $\Lambda$CDM and dark energy with
the same equation of state in the framework of Einstein general
relativity. We find that the perturbation in $f(T)$ gravity grows
slower than that in Einstein general relativity if $\p f/\p T>0$ due
to the effectively weakened gravity. }


\begin{document}

\section{Introduction}
The cause for the late-time accelerated expansion of the universe
remains one of the most compelling problems in modern physics. Many
schemes have been proposed to explain this phenomenon. Although dark
energy scenario \cite{Weinberg:1988cp} is the most popular one among
them, ones have considered some models based on infra-red
modifications to general relativity (GR), such as scalar-tensor
theories \cite{scalar tensor}, f(R) gravity \cite{fR} and braneworld
models \cite{braneworld}. In general, the resulting field equations
are fourth order because the Ricci scalar is constructed from the
second order derivatives of the metric, and this feature may lead to
pathologies. Recently, an alternative model based on modified
teleparallel gravity receives considerable attention. See
\cite{Bengochea:2008gz,Linder,Bamba,Wu,Yerzhanov,Yang,Tsyba,Dent:2010va,Bengochea:2010sg,Wu:2010av,Bamba:2010iw,Karami:2010bu,Karami:2010xy,Dent:2010bp,Ferraro:2006jd}
in detail. Instead of describing gravitational interaction with
curvature of the background spacetime by employing the torsionless
Levi-Civita connection, one can explore the opposite way and resort
to the Weitzenb\"{o}ck connection that has no curvature, in this
case torsion will independently do the job that curvature does in GR
\cite{new GR}. There are some terms in the modified Friedmann
equation that can be identified as the effective dark energy to give
rise to the accelerated expansion of the late-time universe.  This
paradigm boasts the significant advantage that the field equations
are second order, as we shall see in the next section.

In order to discriminate different models, we need to break the
degeneracy of background expansion history, and work in first order
perturbation to find more information concerning different models
\cite{Wang:1998gt,Starobinsky:1998fr,Linder:2003dr,Nojiri:2006ri}.
The matter density perturbation which characterizes the
inhomogeneities of the universe comes to the rescue. While different
models may have the same background behavior, their linear growths
of matter perturbation can be quite different. See, for example,
\cite{fR growth}. In this paper, we derive the evolution equation of
the linear matter density perturbation and find that it takes the
same form as the counterpart in GR, except that the effective
Newton's constant is rescaled by a term related to the first
derivative of $f(T)$. Note that when the Lagrangian of gravitational
part is torsion scalar $T$, it is equivalent to the Einstein-Hilbert
Lagrangian of GR up to a divergence, and hence all behaviors of this
theory reduce to those in GR, including the local Lorentz symmetry.
While for more general $f(T)$ gravity, local Lorentz transformation
fails as a symmetry of this theory \cite{LLT}, and it is expected
that this equation may also provide us some hint about Lorentz
violation.

The outline of this paper is as follows. In Sec.~2, we briefly
review the theoretical structure of teleparallel gravity and how it
explains cosmic acceleration. In Sec.~3, we present the first order
equations based on metric perturbations and vierbein perturbations
respectively. Moreover, we derive the governing equation for matter
density perturbation, then solve it numerically and compare our
result to the counterpart of GR. Finally in Sec.~4, we discuss and
summarize our results.

\section{A brief review of f(T) cosmology}
In Riemann-Cartan spacetime, the curvature tensor and the torsion
tensor coexist. On a manifold, one can define a large number of
connections, which differ from each other up to a tensor quantity.
The assumptions of torsion-free and metric compatibility lead to the
Levi-Civita connection, and this is the one on which Einstein
general relativity is based. However, we are free to choose other
connections, for instance the Weitzenb\"{o}ck connection which is
defined by
\begin{equation}\label{connection}
\Gamma^\lambda_{\mu\nu}=e^\lambda{}_A\partial_\nu e_\mu{}^A,
\end{equation}
where $\mathbf{e}_A(x^\mu)$ is a set of orthonormal vectors, which
form a noncoordinate basis for the tangent space at each point on
this manifold, and $\mathbf{e}^A(x^\mu)$ is the dual vectors. The
torsion tensor is given by
\begin{equation}
T^\lambda{}_{\mu\nu}=\Gamma^\lambda_{\nu\mu}-\Gamma^\lambda_{\mu\nu}=e^\lambda{}_A(\partial_\mu
e_\nu{}^A-\partial_\nu e_\mu{}^A),
\end{equation}
one can find that the curvature tensor and the covariant derivatives
of $\mathbf{e}_A(x^\mu)$ with respect to this connection vanish
globally, therefore $\mathbf{e}_A(x^\mu)$ are absolutely parallel
vector fields, and this theory is dubbed teleparallel gravity
\cite{new GR}. In this formalism, the fundamental dynamical object
is the vierbein field $e_\mu{}^A(x)$, and the metric tensor is
obtained by a byproduct
\begin{equation}
g_{\mu\nu}(x)=\eta_{AB}e_\mu{}^A(x) e_\nu{}^B(x),
\end{equation}
and then the Levi-Civita connection can be defined in a natural way.
The difference between these two connections is described by the
contorsion tensor which takes the form
\begin{equation}
K^{\mu\nu}{}_\rho=-\frac{1}{2}(T^{\mu\nu}{}_\rho-T^{\nu\mu}{}_\rho-T_\rho{}^{\mu\nu}).
\end{equation}
Finally, we can define a torsion scalar as follows
\begin{equation}
T=S_\rho{}^{\mu\nu}T^\rho{}_{\mu\nu},
\end{equation}
where
\begin{equation}
S_\rho{}^{\mu\nu}=\frac{1}{2}(K^{\mu\nu}{}_\rho+\delta^\mu_\rho
T^{\alpha\nu}{}_\alpha-\delta^\nu_\rho T^{\alpha\mu}{}_\alpha).
\end{equation}
$T$ is the simplest teleparallel Lagrangian, which differs
from Einstein-Hilbert Lagrangian only up to a boundary term
\cite{boundary}.

Similar to $f(R)$ gravity, we can write down the Lagrangian for the gravity as a function of $T$. The full action reads
\begin{equation}
I=\frac{1}{2\kappa^2}\int d^4x e\cdot [T+f(T)]+\int d^4x e\cdot
{\cal L}_m, \label{action}
\end{equation}
where $e=\hbox{det}(e_\mu{}^A)=\sqrt{-g}$, $\kappa^2=8\pi G$($G$ is
the Newton's coupling constant), and ${\cal L}_m$ stands for the
matter Lagrangian. Performing variation in this action with respect
to the vierbein yields the equations of motion \m\label{EOM}
-\frac{1}{4}e^\alpha{}_A[T+f]+e^\beta{}_A
T^\mu{}_{\nu\beta}S_\mu{}^{\nu\alpha}[1+f_T]&+&e^{-1}\partial_\mu (e
e^\rho{}_A S_\rho{}^{\mu\alpha})[1+f_T] \nonumber\\
&+&e^\rho{}_A S_\rho{}^{\mu\alpha} f_{TT}\partial_\mu T =
\frac{\kappa^2}{2}e^\rho{}_A \mathbf{T}_\rho{}^\alpha, \n where the
subscript `$T$' denotes the derivative with respect to the torsion
scalar and $\mathbf{T}_\rho{}^\alpha$ is the energy-monmentum
tensor. Comparing to the equation in $f(R)$ gravity which  is fourth
order differential equation, this equation has an advantage of being
second order. So it is much easier for us to analyze it.

From now on, we focus on the a spatially flat Friedmann-Robtson-Walker (FRW)  universe only filled with dust-like matter. The metric is given by
\e
ds^2=dt^2-a^2(t) d{\vec x}^2.
\q
In this case the torsion scalar is related to the Hubble parameter $H\equiv d\ln a/dt$ by
\e
T=-6H^2.
\q
The background equations of motion become \cite{Linder}
\m
H^2&=&{\kappa^2\over 3}\rho-{f\over 6}-2H^2 f_T, \label{effde} \\
\dot H&=& -{1\over 4} {6H^2+f+12 H^2 f_T \over 1+f_T-12 H^2 f_{TT}},
\n where $\rho$ is matter energy density. Evidently, the last two
terms at the right-hand side of Eq.(\ref{effde}) can be explained as
the effective dark energy whose energy density is given by \m
\rho_{de}=\frac{1}{2\kappa^2}(-f+2Tf_T), \label{de} \n and the
corresponding equation of state is
\begin{equation}
w=-1+\frac{(f-T-2Tf_T)(f_T+2Tf_{TT})}{(1+f_T+2Tf_{TT})(f-2Tf_T)}.
\end{equation}
From the above equation, the effective dark energy becomes an
effective cosmological constant if \e f-T-2Tf_T=0, \label{ftft} \q
or \e f_T+2Tf_{TT}=0, \label{ftftt} \q for all $T$. The solution of
Eq.(\ref{ftft}) is \e f(T)=-T+c_1 \sqrt{-T}. \q But now
$(1+f_T+2Tf_{TT})$ equals to zero as well. So this solution does not
provide an effective cosmological constant. Switch to
Eq.(\ref{ftftt}), the solution is \e f(T)=c_2\sqrt{-T}-\kappa^2
\Lambda, \q where $\Lambda$ is a constant. Substituting the above
solution into Eq.(\ref{de}), we find that the effective dark energy
density is simplified to be $\Lambda$ which is nothing but a
cosmological constant and the term $c_2\sqrt{-T}$ does not
contribute to the effective energy density at all. To summarize,
$f(T)$ can be taken as a cosmological constant only when $f(T)$ is a
constant. However it is quite trivial.


\section{Growth factor}

The early universe was made very nearly uniform by an inflationary
state. The origin of structure in the universe is seeded by the
small quantum fluctuations generated at the inflationary epoch.
These small perturbations over time grew to become all of the
structure we observe. Once the universe becomes matter dominated
primeval density inhomogeneities $(\delta\rho/\rho\sim 10^{-5})$ are
amplified by gravity and grow into the structure we see today.In
this section, we investigate how the matter density perturbation
grows in $f(T)$ gravity. We keep terms up to the first order in the
perturbed vierbein field. For the sake of simplicity, we will work
in Newtonian gauge, which is valid for $f(T)$ gravity theory because
it still preserves the principle of general covariance.

This section
is divided into three subsections. Firstly, we follow the approaches
in \cite{Dent:2010bp} and define all the scalar degrees of freedom
in the perturbed metric in Newtonian gauge. We find that this ansatz is too naive and problematic. In the second subsection, we
start with a general perturbed vierbein field which includes more
degrees of freedom and derive the correct evolution equation for the matter density perturbation. In the last
subsection, we consider a concrete $f(T)$ model and compare the growth of matter over-dense perturbation in $f(T)$ gravity with that in GR.

\subsection{A naive ansatz for the perturbed vierbein}

Up to the linear order, scalar perturbations should decouple with vector
and tensor perturbations. The perturbed FRW metric can be written by
\begin{equation}
ds^2=(1+2\phi)dt^2-a^2(t)(1-2\psi)\delta_{ij}dx^i dx^j.
\end{equation}
A naive ansatz for the perturbed vierbein can be written by
\begin{equation}
e_\mu{}^A=\left(
\begin{array}{cc}
1+\phi & 0 \\
0 & a(1-\psi)\delta_i{}^m
\end{array}\right).
\end{equation}
Accordingly, the perturbed energy-momentum tensor takes the form
\begin{equation}
\delta\mathbf{T}_\mu{}^\nu=\left(
\begin{array}{cc}
-\delta \rho & -a^{-2}(\rho+p)\partial_i v \\
(\rho+p)\partial_i v & \delta_i^j\delta p
\end{array} \right),
\end{equation}
where $\rho$ is the energy density, $p$ is the pressure and $v$ is
the velocity potential. From the action in (\ref{action}), the first
order perturbations are governed by
\begin{eqnarray}
&&E^0{}_0:\ \frac{\kappa^2}{2}\delta\rho=-\frac{k^2}{a^2}\psi(1+f_T)-3H(\dot{\psi}+H\phi)(1+f_T-12H^2f_{TT}) \label{delta rho1},\\
&&E^i{}_0:\ \frac{\kappa^2}{2}(\rho+p)\partial^i v=-(\partial^i \dot{\psi}+H\partial^i \phi)(1+f_T)+12H\dot{H}f_{TT}\partial^i \psi \label{v11},\\
&&E^0{}_i:\ \frac{\kappa^2}{2}(\rho+p)\partial_i v=-(\partial_i \dot{\psi}+H\partial_i \phi)(1+f_T-12H^2f_{TT})\label{v21},\\
&&E^i{}_j(i=j):\nonumber\\
&&\ \ \ \ \ \ \ \ \frac{\kappa^2}{2}\delta p=-(36H^4\phi+60H^2\dot{H}\phi+12H^3\dot{\phi}+36H^3\dot{\psi}+36H\dot{H}\dot{\psi}+12H^2\ddot{\psi})f_{TT}\nonumber\\
&&\ \ \ \ \ \ \ \ \ \ \ \ \ \ +(3H^2\phi+H\dot{\phi}+2\dot{H}\phi+3H\dot{\psi}+\ddot{\psi})(1+f_T)+144H^3\dot{H}f_{TTT}(\dot{\psi}+H\phi),\label{delta p1}\\
&&E^i{}_j(i \neq j):\ \psi-\phi =0,
\end{eqnarray}
which are the same as those in \cite{Dent:2010va,Dent:2010bp}, and
we also use $E^\mu{}_A$ to denote the equation obtained from
variation of the action with respect to $e_\mu{}^A$. Note that
$\partial^i=\delta^{ij}\partial_j$ and
$\partial^2=\partial^i\partial_i$ throughout this work. Comparing
Eq.\eqref{v11} with Eq.\eqref{v21}, we approach to an extra
scale-independent constraint on $\phi$, namely
\begin{equation}
\dot{H}\partial_i\psi=H(\partial_i\dot{\psi}+H\partial_i\phi),
\end{equation}
if $f_{TT}\neq 0$. Or equivalently, there are the same number of
degrees of freedom as that in GR, but $f(T)$ theory leads to one
more equation. It may lead to inconsistency. Besides, the
scale-independent evolution of $\phi$ in the above equation is
incompatible with the the integrated Sachs-Wolfe effect.

One may consider $f_{TT}=0$. However, if so, $f(T)\propto T$ and $f(T)$ gravity is effectively reduced to GR. In order to solve this puzzle, we go to a more general ansatz in the next subsection.

\subsection{A general ansatz on perturbed vierbein in $f(T)$ gravity}

Since the local Lorentz symmetry is broken down in $f(T)$ gravity,
extra degrees of freedom compared to GR should appear. In this
subsection, we include all scalar degrees of freedom in the
vierbein. Following \cite{hamilton}, the perturbed vierbein is
expressed in terms of unperturbed vierbein $\bar{e}_\mu{}^A$ and a
first-order quantity $\chi_A{}^B$ as
\begin{equation}\label{e}
e_\mu{}^A=(\delta_B^A+\chi_B{}^A)\bar{e}_\mu{}^B,
\end{equation}
with $\bar{e}_0{}^A=\delta_0^A$ and $\bar{e}_i{}^A=a\delta_i^A$. In
this subsection, we only take into account the scalar degrees of freedom,
which are encoded in $\chi_B{}^A$ as follows
\begin{equation}\label{chi}
\chi_{AB}=\left(
\begin{array}{cc}
\phi & \partial_i w\\
\partial_i \tilde{w} & \ \ \ \ \ \delta_{ij}\psi+\partial_i\partial_j
h+\epsilon_{ijk}\partial^k \tilde{h}
\end{array}\right).
\end{equation}
Keep in mind that the captical indices are lowered or raised by the
Minkowski metric $\eta_{AB}$ or its inverse. There are six scalar
degrees of freedom altogether. The corresponding perturbed
metric takes the form:
\begin{equation}
g_{\mu\nu}=\left(
\begin{array}{cc}
1+2\phi & a\partial_i(w+\tilde{w})\\
a\partial_i(w+\tilde{w}) & \ \ \ \
-a^2\((1-2\psi)\delta_{ij}-2\partial_i\partial_j h\)
\end{array}\right).
\label{metricc}
\end{equation}
In general, $w$ and $\tilde w$ always affect the metric in terms of
their combination $w+\tilde w$, which serves as a single degree of freedom, but as
far as the vierbein is concerned, they are two independent degrees of freedom.
Moreover, $\tilde{h}$ does not present itself in the metric as well. To summarize,
there are two degrees of freedom which do not contribute to the metric, and this
is exactly what we ignore in the previous subsection.

In longitudinal gauge, $\tilde{w}=-w$ and $h=0$. Since $w$ has a
mass dimension, we introduce a dimensionless quantity $\zeta$ which
is related to $w$ by $\zeta=aHw$. As demonstrated in the appendix,
one can obtain the perturbed equations up to first order as follows
\begin{eqnarray}
&&E^0{}_0:\ \frac{\kappa^2}{2}\delta\rho=a^{-2}(1+f_T)\partial^2\psi-12a^{-2}H^2f_{TT}\partial^2\zeta-3H(1+f_T-12H^2f_{TT})(\dot{\psi}+H\phi), \ \ \ \ \ \ \ \ \ \ \label{delta rho2}\\
&&E^i{}_0:\ \frac{\kappa^2}{2}(\rho+p)\partial^i v
=-(1+f_T)(\partial^i
\dot{\psi}+H\partial^i \phi)+12H\dot{H}f_{TT}\partial^i \psi,\label{v12}\\
&&E^0{}_i:\ \frac{\kappa^2}{2}(\rho+p)\partial_i v =-(1+f_T-12H^2f_{TT})(\partial_i \dot{\psi}+H\partial_i \phi)-4a^{-2}H f_{TT}\partial_i\partial^2\zeta,\label{v22}\\
&&\hbox{Tr}(E^i{}_j):\nonumber\\
&&\ \ \ \ \ \ \ \ \ \frac{\kappa^2}{2}\delta p
=(1+f_T)\((\ddot{\psi}+3H\dot{\psi}+H\dot{\phi}+2\dot{H}\phi+3H^2\phi)-\frac{1}{3}
a^{-2}\partial^2(\psi-\phi)\)\nonumber\\
&&\ \ \
\ \ \ \ \ \ \ \ \ \ \ \ \ \ \ \ +f_{TT}\(-12H^2\ddot{\psi}-36H(\dot{H}+H^2)\dot\psi-12H^3\dot\phi-(60\dot{H}H^2+36H^4)\phi\right.\nonumber\\
&&\ \ \ \ \ \ \ \ \ \ \ \ \ \ \ \ \ \ \ \ \ \ \ \ \ \
\left.+a^{-2}(8\dot{H}+4H^2)\partial^2 \zeta+4a^{-2}H\partial^2\dot
\zeta\)\nonumber\\
&&\ \ \ \ \ \ \ \ \ \ \ \ \ \ \ \ \ \
+12f_{TTT}\dot{H}H^2\(12H(\dot\psi+H\phi)-4a^{-2}\partial^2 \zeta\)\label{delta p2},\\
&&E^i{}_j(i \neq j):\
(1+f_T)\partial_i\partial^j(\phi-\psi)+12\dot{H}f_{TT}\partial_i\partial^j
\zeta=0\label{the fifth}.
\end{eqnarray}
See Appendix \ref{ap} in detail. 
Note that in the above equations, the Parity-violating term $\tilde h$
disappears, but $w$ survives, even though $w$ does not appear in the perturbed metric in longitudinal gauge. Compared to the first order
equations in GR \cite{mukhanov}, we have an extra degree of freedom
$\zeta$ in the perturbed equations and one more equation is obtained. Our equations are self-consistent. On the other hand,  for the trivial
Lagrangian in which $f(T)$ is a linear function of the torsion scalar $T$, all
analysis should parallel to those in GR except for a re-scaled
coupling constant, and the extra degree of freedom $\zeta$ should disappear. Here we see that $\zeta$ always appears in the company of $f_{TT}$. It is quite reasonable.

Since the matter Lagrangian is invariant under general coordinate
transformation, the energy-momentum tensor should be conserved with
respect to the Levi-Civita connection. Then one can find two
equations which take exactly the same form as their counterparts in
GR:
\begin{eqnarray}
&&\dot{\delta \rho}+3H(\delta \rho+\delta
p)+a^{-2}(\rho+p)\partial^2
v-3(\rho+p)\dot{\psi}=0\label{dot deltarho2}, \\
&&\dot{p}\partial_i v+(\rho+p)\partial_i \dot{v}+\partial_i \delta
p+(\rho+p)\partial_i \phi=0\label{dot v}.\\
&&\frac{d}{dt}\left((\rho+p)\partial_i v\right)+3H(\rho+p)\partial_i
v+\partial_i p+(\rho+p)\partial_i \phi=0.
\end{eqnarray}
One can also derive the above two equations from Eq.(\ref{delta rho2} -\ref{the fifth}). From now on we will focus on the universe only filled
with dust-like matter, namely $p=\delta p=0$, and all equations will
be transformed to Fourier space. From Eq.\eqref{dot v}, one reaches
a very useful relation
\begin{equation}\label{vphi}
\dot{v}=-\phi.
\end{equation}
Define a gauge invariant fractional matter perturbation
\begin{equation}\label{delta def}
\delta_m\equiv \frac{\tilde{\delta \rho_m}}{\rho_m},
\end{equation}
where \m \tilde{\delta \rho_m}\equiv \delta \rho_m-3H\rho_m v \n is
the gauge-invariant comoving matter density perturbation. It can
also be interpreted as the density perturbation on spacelike
hypersurfaces orthogonal to comoving worldlines. Considering
Eq.\eqref{delta rho2} and Eq.\eqref{v22}, we obtain \e
 \frac{\kappa^2}{2}\tilde{\delta \rho_m}=-\frac{k^2}{a^2}\psi(1+f_T).
\label{deltam} \q From Eqs. \eqref{v12}, \eqref{the fifth} and
\eqref{delta def}, the evolution of matter density perturbation is
given by
\begin{equation}\label{dot delta}
\dot\delta_m=\ka v-12H\dot{H}f_{TT}\ka\frac{\zeta}{\frac{\kappa^2}{2}\rho_m}.
\end{equation}
In order to get the evolution equation of $\delta_m$, we need to work out the solution of $\zeta$ as well.

In this subsection we focus on the non-trivial case with $f_{TT}\neq 0$.  From Eqs. \eqref{v12} and \eqref{v22},  we obtain
\e
3\dot{H}\psi=3H\dot{\psi}+3H^2\phi+\frac{k^2}{a^2}\zeta.
\label{ppz}
\q
On the other hand, Eq. \eqref{the fifth} can be written by
\e
\phi=\psi-12\dot H {f_{TT}\over 1+f_T}\zeta.
\q
Combing the above two equations, we find
\begin{eqnarray}
\zeta&&=\frac{3(\dot H-H^2)\psi-3H \dot\psi}{\frac{k^2}{a^2}-36\dot H H^2{f_{TT}\over 1+f_T}}.
\label{zeta}
\end{eqnarray}
In the subhorizon limit, $\zeta\sim {a^2H^2\over k^2}\psi\ll \psi$. One can expect that $\zeta$ will play an important role on the evolution of perturbations at large scales. But here we only focus on the physics in the subhorizon limit and we have
\begin{equation}\label{phipsi}
\phi\simeq \psi.
\end{equation}
It is the same as that in minimally coupled GR. Therefore, from Eq.
\eqref{v12}, $Hv\sim \phi\simeq \psi\gg \zeta$ and hence the second
term on the right-hand side of Eq. \eqref{dot delta} can be
neglected. Taking the time derivative of Eq. \eqref{dot delta}, one
can obtain
\begin{equation}\label{ddot delta}
\ddot \delta_m+2H\dot \delta_m+{k^2\over a^2}\phi= 0,
\end{equation}
where Eq.\eqref{vphi} is taken into account. Plugging
Eq.\eqref{phipsi} into this equation and combining with
Eq.\eqref{deltam}, the evolution equation of linear matter
perturbation becomes
\begin{equation}\label{delta}
\ddot\delta_m+2H\dot\delta_m-4\pi G_{\text{eff}}\rho_m \delta_m=0,
\end{equation}
where $G_{\text{eff}}$ is the effective Newton's constant which is
related to $G$ by \e G_{\text{eff}}={G\over 1+f_T}. \q If $f(T)$ is
a linear function of $T$, namely $f(T)=\alpha T$, the Newton's
constant is just rescaled to be $G_{\text{eff}}=G/(1+\alpha)$, which
is also constant in time. We can understand this result from the
action (\ref{action}) directly. In $f(R)$ or scalar-tensor gravity,
one can get an analogous equation for fractional matter perturbation
$\delta_m$ with a redefinition of Newton's constant in the short
wave-length limit \cite{fR growth}.

At the deep matter dominant era, if $f_T\simeq 0$, the solution of Eq. (\ref{delta}) indicates that the matter density perturbation goes like $\delta_m\propto a$. Therefore $\tilde{\delta \rho_m}=\rho_m \delta_m\sim a^{-2}$ and $\phi\simeq \psi\sim$ constant in time. From Eq.(\ref{ppz}), $\zeta\sim a^{-1}$ which implies that $\zeta$ decreases with respect to $\phi$ and $\psi$.

To summarize, we consider all the scalar degrees of freedom in this
subsection and obtain the evolution equation of matter energy density perturbation in the subhorizon limit. In this limit, the extra degree of freedom $\zeta$ is suppressed compared to $\phi$ and $\psi$, but it removes the inconsistency in the former subsection.


\subsection{Numerical analysis}
For convenience, we define the growth as the ratio of the
perturbation amplitude at some scale factor relative to some initial
scale factor, $D=\delta_m (a)/\delta_m(a_i)$. The matter density
perturbation $\delta_m$ is proportional to the scale factor $a$ in
the $f(T)$ gravity with $f_T\ll 1$ during the matter era. We
introduce a new variable $g(a)$, namely \e g(a)\equiv {D(a)\over a}
\q which does not depend on $a$ during the matter era; thus, the
natural choice for the initial conditions are $g(a_{i})=1, dg/d\ln
a|_{a=a_i}=0$. From Eq.(\ref{delta}), the equation for $g(a)$
becomes
\begin{equation}
\frac{d^2 g}{d \ln a^2}+\(4+\frac{\dot{H}}{H^2}\)\frac{d g}{d \ln
a}+\(3+\frac{\dot{H}}{H^2}-\frac{4\pi G_{\text{eff}} \rho_m}{H^2}\)
g=0.
\end{equation}
This equation reduces to that for dark energy scenario in GR with
the same equation of state as the effective dark energy in $f(T)$ if
we replace $G_{\text{eff}}$ by $G$.  For a universe only filled with
dust-like matter in $f(T)$ gravity, we have \m\label{H dot} {\dot
H\over H^2}=-{3\over 2}{1+f/6H^2+2f_T \over 1+f_T-12H^2 f_{TT}}, \n
and \m\label{eff G} \frac{4\pi G_{\text{eff}}
\rho_m}{H^2}={3\Omega_{m}^0\over 2(1+f_T)}{H_0^2\over H^2} a^{-3},
\n here the scale factor $a_0$ is normalized to be one and
$\Omega_m^0$ is the matter energy density parameter. The superscript
`0' denotes that the variables are evaluated at present.

For intance, we consider a power law model with
\begin{equation}
f(T)=\alpha(-T)^n=\alpha(6H^2)^n,
\end{equation}
where $\alpha$ can be determined by the present Hubble parameter and matter density parameter, namely
\e
\alpha=(6H^2_0)^{1-n}(1-\Omega_m^0)/(2n-1).
\q
The equation for the perturbation becomes
\m
\frac{d^2 g}{d \ln
a^2}&+&\[4-\frac{3}{2}\frac{1-h^{2n-2}(1-\Omega_m^0)}{1-n
h^{2n-2}(1-\Omega_m^0)}\]\frac{d g}{d \ln
a} \nonumber \\
&+&\[3-\frac{3}{2}\frac{1-h^{2n-2}(1-\Omega_m^0)}{1-n
h^{2n-2}(1-\Omega_m^0)}-\frac{3\Omega_m^0
h^{-2}a^{-3}}{2(1-\frac{n(1-\Omega_m^0)}{2n-1}h^{2n-2})}\] g=0, \label{g}
\n
where $h\equiv H/H_0$ which is governed by
\begin{equation}\label{h}
\frac{d h^2}{d \ln a}=\frac{-3h^2+3h^{2n}(1-\Omega_m^0)}{1-n
h^{2n-2}(1-\Omega_m^0)}.
\end{equation}
Since the above equations are too complicated to be solved
analytically, we will use numerical method. In \cite{Linder}, Linder
pointed out that this model can fit current observation only when
$n\ll 1$. Therefore we have \e f_T={n\over
1-2n}(1-\Omega_m^0)\({H_0\over H}\)^{2(1-n)} \q which is much
smaller than one during the matter era because $n\ll 1$ and
$H_0/H\ll 1$. Here we will adopt $n=0.1$ and $\Omega_m^0=0.28$ for
numerical calculation. The initial moment should be taken during the
matter era, e.g., $a_i=1/31$ (i.e., $z=30$). In addition, the
initial condition of $h(a)$ is $h(a=1)=1$.

\begin{figure}[h]
\begin{center}
\includegraphics[width=12cm]{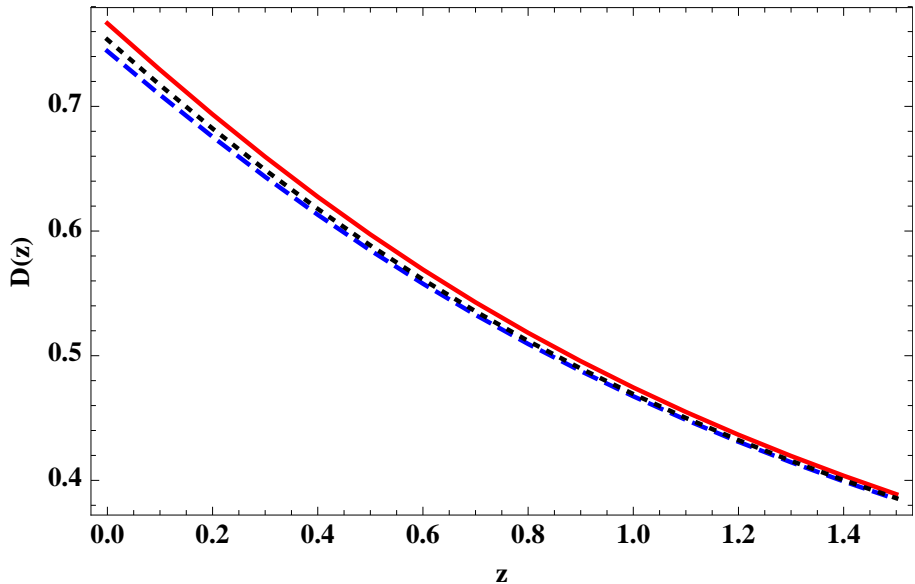}
\end{center}
\caption{The numerical soluion of the evolution equation for the
growth rate of matter perturbation in $\Lambda$CDM(the red solid
line), power law f(T) model(the blue dashed line) and dark energy
with the same equation of state in GR (the black dotted line).  }
\label{fig:dz}
\end{figure}

When $f$ is a constant or $n=0$, the term $f(T)$ acts just as a
cosmological constant. In this case, our result recovers that in
$\Lambda$CDM in GR. For a universe filled with matter and dark
energy whose equation of state is the same as that in $f(T)$
gravity, the evolution equation of $\delta_m$ in the framework of GR
can be obtained by replacing $G_{\text{eff}}$ with $G$ in
(\ref{delta}). Our numerical results are illustrated in
Fig.~\ref{fig:dz}.
Since $f_T>0$, the effective Newton's constant in $f(T)$ gravity gets smaller than that in GR, and the gravitational interaction is weakened. That is why the
over-dense perturbation in $f(T)$ gravity grows slower than that in GR (See the blue dashed and black dotted lines).

\section{Discussions}

In this paper we derived the evolution equation for linear matter density
perturbation in the framework of $f(T)$ gravity and compared it to
that in GR. We began our analysis from two aspects. One is
based on a naive ansatz in which we chose the simplest vierbein and concluded that it leads to an inconsistency in Sec.~3.1. Though different vierbeins are related to each
other by a Lorentz transformation, they may have different
predictions because the Lorentz symmetry is broken down in $f(T)$ theory. However, there is not a principle for us to choose a `physical' vierbein. In Sec.~3.2, we proposed a strategy to solve this puzzle. We started with the most general  perturbed vierbein and we found that an extra degree of freedom cure  the inconsistency in the former case even though it does not appear in the perturbed metric. Finally, in Sec.~3.3, we figured out the growth factor in the power law $f(T)$ model in detail and showed that the over-dense matter perturbation grows slower than that in GR due to the weakened gravity.


When $f(T)$ contains some nonlinear terms of $T$, $f(T)$ gravity is not Lorentz invariant any more. However, the Lorentz symmetry should be preserved at least at small scales. This may require a stringent constraint on $f(T)$ gravity. Once we take this constraint into account, whether $f(T)$ theory can lead to an accelerated expansion of our universe is still an open question. We will come back to this question in the future.

\vspace{1.0cm}

\noindent {\bf Acknowledgments}

\vspace{.5cm}

RZ would like to thank M.~Li and E.~Linder for helpful discussions.
This work is supported by the project of Knowledge Innovation
Program of Chinese Academy of Science and a grant from NSFC.

\vspace{1.5cm}

\appendix

\section{Equations of motion for the perturbations}
\label{ap}

In the appendix, we present some details of our calculations. From
the definition in Eqs. \eqref{e} and \eqref{chi}, one obtains the
components of perturbed vierbein $e_\mu{}^A$:
\begin{equation*}
\begin{array}{ll}
e_0{}^0=1+\phi, &\ \ \ \ \ \ e_i{}^0=a\partial_i\tilde{w},\\
e_0{}^i=-\partial^i w, &\ \ \ \ \ \
e_j{}^i=a\((1-\psi)\delta_j^i-\partial_j\partial^i
h-\epsilon_j{}^{in}\partial_n\tilde{h}\),
\end{array}
\end{equation*}
and its inverse $e^\mu{}_A$:
\begin{equation*}
\begin{array}{ll}
e^0{}_0=1-\phi, &\ \ \ \ \ \ e^i{}_0=a^{-1}\partial^i w,\\
e^0{}_i=-\partial_i \tilde{w}, &\ \ \ \ \ \
e^i{}_j=a^{-1}\((1+\psi)\delta_j^i+\partial_j\partial^i
h+\epsilon_j{}^{in}\partial_n\tilde{h}\).
\end{array}
\end{equation*}
Plugging the above equations into $g_{\mu\nu}=\eta_{AB}e_\mu{}^A
e_\nu{}^B$, one gets the metric in Eq. \eqref{metricc}. For
simplicity, we choose the Newtonian gauge, where $\tilde w=-w$, and
$h=0$. In this case,
\begin{eqnarray*}
&&T^0{}_{0i}=-\partial_i\phi-a\partial_i \dot w,\\
&&T^0{}_{ij}=0,\\
&&T^i{}_{0j}=H\delta^i_j-\dot{\psi}\delta^i_j-\epsilon_j{}^{in}\partial_n\dot{\tilde{h}}+a^{-1}\partial_j\partial^i
w,\\
&&T^i{}_{jk}=\partial_k \psi
\delta^i_j-\partial_j\psi\delta^i_k+\epsilon_j{}^{in}\partial_k\partial_n\tilde
h-\epsilon_k{}^{in}\partial_j\partial_n\tilde h,
\end{eqnarray*}
and then
\begin{eqnarray*}
&&S_0{}^{0i}=a^{-2}\partial^i\psi,\\
&&S_0{}^{ij}=-\half a^{-2}\epsilon^{ijn}\partial_n\dot{\tilde h},\\
&&S_i{}^{0j}=-H\delta_i^j+(\dot\psi+2H\phi-\half a^{-1}\partial^2
w)\delta_i^j+\half a^{-1}\partial_i\partial^j w,\\
&&S_i{}^{jk}=-\half
a^{-2}\epsilon^{jkn}\partial_i\partial_n\tilde{h}+\half
a^{-2}(\partial^k\psi-\partial^k\phi-a\partial^k
\dot{w})\delta^j_i-\half
a^{-2}(\partial^j\psi-\partial^j\phi-a\partial^j \dot{w})\delta^k_i.
\end{eqnarray*}
In addition, one can also calcualte perturbed torsion scalar, namely
\e
T=-6H^2+12H(\dot \psi+H\phi)-4a^{-1}H\partial^2 w.
\q

Defining $\tilde{T}^\alpha{}_A=e^\rho{}_A \mathbf{T}_\rho{}^\alpha$, one can easily obtain
\begin{eqnarray}
&&\tilde{T}^0{}_0=-\rho-\delta\rho+\rho\phi,\\
&&\tilde{T}^0{}_i=a^{-1}(\rho+p)\partial_i v-\rho\partial_i w,\\
&&\tilde{T}^i{}_0=-a^{-2}(\rho+p)\partial^i v+a^{-1}p\partial^i w,\\
&&\tilde{T}^i{}_j=a^{-1}(p+\delta
p)\delta^i_j+a^{-1}p\psi\delta^i_j+a^{-1}p\epsilon_j{}^{in}\partial_n\tilde
h.
\end{eqnarray}
Pluging these expressions into the equation of motion Eq. \eqref{EOM}, we find\\
$E^0{}_0$:\ \\
\e
\frac{\kappa^2}{2}\rho=\frac{1}{4}(T+f)+3H^2(1+f_T),
\label{bc rho}
\q
and
\m
\frac{\kappa^2}{2}(-\delta\rho+\rho\phi)&=&\frac{1}{4}(T+f)\phi+(1+f_T)(3H\dot{\psi}+6H^2\phi-a^{-2}\partial^2\psi)\nonumber \\
&-&3H^2f_{TT}(12H(\dot{\psi}+H\phi)-4a^{-1}H\partial^2 w).
\label{pt rho}
\n
Note that Eq. \eqref{bc rho} describes evolution of homogeneous background and Eq.\eqref{pt rho} equation of motion for density perturbation $\delta \rho$. Combing Eq. \eqref{bc rho} and Eq. \eqref{pt rho} yields Eq. \eqref{delta rho2}.\\
$E^0{}_i$:\ \\
\begin{eqnarray}
\frac{\kappa^2}{2}\(a^{-1}(\rho+p)\partial_i v-\rho\partial_i
w\)&=&-\frac{1}{4}(T+f)\partial_i w+(1+f_T)(-a^{-1}\partial_i
\dot{\psi}-a^{-1}H\partial_i \phi-3H^2\partial_i
w)\nonumber\\
&+&f_{TT}(12a^{-1}H^2(\partial_i\dot\psi+H\partial_i\phi)-4a^{-2}H^2\partial_i\partial^2 w).
\end{eqnarray}
Considering Eq. \eqref{bc rho}, one can obtain Eq. \eqref{v12}.\\
$E^i{}_0$:\\
\begin{eqnarray}
\frac{\kappa^2}{2}\(-a^{-2}(\rho+p)\partial^i v+a^{-1}p\partial^i
w\)&=&-\frac{1}{4}a^{-1}(T+f)\partial^i
w-12a^{-2}H\dot{H}f_{TT}(\partial^i \psi-a H\partial^i w)
\nonumber\\
&+&(1+f_T)\(a^{-2}\partial^i(
\dot{\psi}+H\phi)-a^{-1}(\dot{H}+3H^2)\partial^i w\),
\label{eio}
\end{eqnarray} \\
$E^i{}_j$: \\
\e
\frac{\kappa^2}{2}p=-\frac{1}{4}(T+f)-(\dot{H}+3H^2)(1+f_T)+12\dot{H}H^2 f_{TT},
\label{bc p}
\q
\begin{eqnarray}
&&\frac{\kappa^2}{2}(\delta
p\delta^i_j+p\psi\delta^i_j+p\epsilon_j{}^{in}\partial_n\tilde
h)\nonumber\\
&&=-\frac{1}{4}(T+f)(\psi\delta^i_j+\epsilon_j{}^{in}\partial_n\tilde{h})+(1+f_T)\((\ddot{\psi}+3H\dot{\psi}-\dot{H}\psi-3H^2\psi+H\dot{\phi}+2\dot{H}\phi+3H^2\phi)\delta^i_j\right.\nonumber\\
&&\ \ \
\left.-(\dot{H}+3H^2)\epsilon_j{}^{in}\partial_n\tilde{h}-\half
a^{-2}\partial^2(\psi-\phi)\delta^i_j+\half
a^{-2}\partial_j\partial^i(\psi-\phi)\)\nonumber\\
&&\ \ \
+f_{TT}\((-12H^2\ddot{\psi}-36H(\dot{H}+H^2)\dot\psi+12\dot{H}H^2\psi-12H^3\dot\phi-(60\dot{H}H^2+36H^4)\phi)\delta^i_j\right.\nonumber\\
&&\ \ \ \ \ \ \ \ \ \left.+a^{-1}(14\dot{H}H\partial^2
w+8H^3\partial^2 w+4H^2\partial^2\dot
w)\delta^i_j+12\dot{H}H^2\epsilon_j{}^{in}\partial_n\tilde{h}-6a^{-1}\dot{H}H\partial_j\partial^i
w\)\nonumber\\
&&\ \ \ \ \
+12f_{TTT}\dot{H}H^2\(12H(\dot\psi+H\phi)-4a^{-1}H\partial^2
w\)\delta^i_j.
\label{delta p0}
\end{eqnarray}
Plugging Eq. \eqref{bc p} into \eqref{eio}, one reaches
Eq.\eqref{v22}. Using Eq. \eqref{bc p}, one can simplify
Eq.\eqref{delta p0} as
\begin{eqnarray}
&&\frac{\kappa^2}{2}\delta
p\delta^i_j\nonumber\\
&&=(1+f_T)\((\ddot{\psi}+3H\dot{\psi}+H\dot{\phi}+2\dot{H}\phi+3H^2\phi)\delta^i_j-\half
a^{-2}\partial^2(\psi-\phi)\delta^i_j+\half
a^{-2}\partial_j\partial^i(\psi-\phi)\)\nonumber\\
&&\ \ \
+f_{TT}\((-12H^2\ddot{\psi}-36H(\dot{H}+H^2)\dot\psi-12H^3\dot\phi-(60\dot{H}H^2+36H^4)\phi)\delta^i_j\right.\nonumber\\
&&\ \ \ \ \ \ \ \ \ \left.+a^{-1}(14\dot{H}H\partial^2
w+8H^3\partial^2 w+4H^2\partial^2\dot
w)\delta^i_j-6a^{-1}\dot{H}H\partial_j\partial^i
w\)\nonumber\\
&&\ \ \ \ \
+12f_{TTT}\dot{H}H^2\(12H(\dot\psi+H\phi)-4a^{-1}H\partial^2
w\)\delta^i_j. \label{delta p01}
\end{eqnarray}
Taking the trace of Eq.\eqref{delta p01}, one obtains
\begin{eqnarray}
\frac{\kappa^2}{2}\delta p
&&=(1+f_T)\((\ddot{\psi}+3H\dot{\psi}+H\dot{\phi}+2\dot{H}\phi+3H^2\phi)-\frac{1}{3}
a^{-2}\partial^2(\psi-\phi)\)\nonumber\\
&&\ \ \
+f_{TT}\(-12H^2\ddot{\psi}-36H(\dot{H}+H^2)\dot\psi-12H^3\dot\phi-(60\dot{H}H^2+36H^4)\phi\right.\nonumber\\
&&\ \ \ \ \ \ \ \ \ \left.+a^{-1}(12\dot{H}H\partial^2
w+8H^3\partial^2 w+4H^2\partial^2\dot
w)\)\nonumber\\
&&\ \ \ \ \
+12f_{TTT}\dot{H}H^2\(12H(\dot\psi+H\phi)-4a^{-1}H\partial^2 w\).
\label{delta p02}
\end{eqnarray}
Noticing that $\zeta=a H w$, we can rewrite this equation in terms
of Eq.\eqref{delta p2}. In addition, by combining Eq.\eqref{delta
p01} and Eq.\eqref{delta p02}, one can obtain Eq.\eqref{the fifth}
after expressing $w$ in terms of the dimensionless quantity $\zeta$.
Here we want to stress that the terms with $\tilde h$ in
\eqref{delta p0} are cancelled and thus $\tilde h$ does not show up
in the perturbation equations in Sec.~3.2.



\begin{thebibliography}{999}



\bibitem{Weinberg:1988cp}
  S.~Weinberg,
  ``The cosmological constant problem,''
  Rev.\ Mod.\ Phys.\  {\bf 61}, 1 (1989); \\
  S.~M.~Carroll,
  ``The cosmological constant,''
  Living Rev.\ Rel.\  {\bf 4}, 1 (2001)
  [arXiv:astro-ph/0004075];\\
  V.~Sahni and A.~A.~Starobinsky,
  ``The Case for a Positive Cosmological Lambda-term,''
  Int.\ J.\ Mod.\ Phys.\  D {\bf 9}, 373 (2000)
  [arXiv:astro-ph/9904398].


\bibitem{scalar tensor}
J.~P.~Uzan,
  ``Cosmological scaling solutions of non-minimally coupled scalar fields,''
  Phys.\ Rev.\  D {\bf 59}, 123510 (1999)
  [arXiv:gr-qc/9903004]; \\
  T.~Chiba,
  ``Quintessence, the gravitational constant, and gravity,''
  Phys.\ Rev.\  D {\bf 60}, 083508 (1999)
  [arXiv:gr-qc/9903094];\\
  B.~Boisseau, G.~Esposito-Farese, D.~Polarski and A.~A.~Starobinsky,
  ``Reconstruction of a scalar-tensor theory of gravity in an accelerating
  universe,''
  Phys.\ Rev.\ Lett.\  {\bf 85}, 2236 (2000)
  [arXiv:gr-qc/0001066].


\bibitem{fR}
  S.~M.~Carroll, V.~Duvvuri, M.~Trodden and M.~S.~Turner,
  ``Is Cosmic Speed-Up Due to New Gravitational Physics?,''
  Phys.\ Rev.\  D {\bf 70}, 043528 (2004)
  [arXiv:astro-ph/0306438];\\
  S.~Nojiri and S.~D.~Odintsov,
  ``Modified gravity with negative and positive powers of the curvature:
  Unification of the inflation and of the cosmic acceleration,''
  Phys.\ Rev.\  D {\bf 68}, 123512 (2003)
  [arXiv:hep-th/0307288];

\bibitem{braneworld}
  V.~Sahni and Y.~Shtanov,``Braneworld models of dark energy", JCAP
  {\bf 11}, 014 (2003)


\bibitem{Bengochea:2008gz}
  G.~R.~Bengochea and R.~Ferraro,
  ``Dark torsion as the cosmic speed-up,''
  Phys.\ Rev.\  D {\bf 79}, 124019 (2009)
  [arXiv:0812.1205 [astro-ph]].

\bibitem{Linder}
  E.~V.~Linder,
  ``Einstein's Other Gravity and the Acceleration of the Universe,''
  Phys.\ Rev.\  D {\bf 81}, 127301 (2010)
  [arXiv:1005.3039 [astro-ph.CO]].

\bibitem{Bamba}
  K.~Bamba, C.~Q.~Geng and C.~C.~Lee,
  ``Cosmological evolution in exponential gravity,''
  JCAP {\bf 1008}, 021 (2010)
  [arXiv:1005.4574 [astro-ph.CO]].

\bibitem{Wu}
  P.~Wu and H.~W.~Yu,
  ``Observational constraints on $f(T)$ theory,''
  Phys.\ Lett.\  B {\bf 693}, 415 (2010)
  [arXiv:1006.0674 [gr-qc]]; \\
  P.~Wu and H.~W.~Yu,
  ``The dynamical behavior of $f(T)$ theory,''
  Phys.\ Lett.\  B {\bf 692}, 176 (2010)
  [arXiv:1007.2348 [astro-ph.CO]].


\bibitem{Yerzhanov}
  K.~K.~Yerzhanov, S.~R.~Myrzakul, I.~I.~Kulnazarov and R.~Myrzakulov,
  ``Accelerating cosmology in F(T) gravity with scalar field,''
  arXiv:1006.3879 [gr-qc];\\
  R.~Myrzakulov,
  ``Accelerating universe from F(T) gravities,''
  arXiv:1006.1120 [gr-qc];\\
  P.~Y.~Tsyba, I.~I.~Kulnazarov, K.~K.~Yerzhanov and R.~Myrzakulov,
  ``Pure kinetic k-essence as the cosmic speed-up and $F(T)$ - gravity,''
  arXiv:1008.0779 [astro-ph.CO];\\
  R.~Myrzakulov,
  ``F(T) gravity and k-essence,''
  arXiv:1008.4486 [astro-ph.CO].




\bibitem{Yang}
  R.~J.~Yang,
  ``New types of $f(T)$ gravities,''
  arXiv:1007.3571 [gr-qc].

\bibitem{Tsyba}
  P.~Y.~Tsyba, I.~I.~Kulnazarov, K.~K.~Yerzhanov and R.~Myrzakulov,
  ``Pure kinetic k-essence as the cosmic speed-up and $F(T)$ - gravity,''
  arXiv:1008.0779 [astro-ph.CO].

\bibitem{Dent:2010va}
  J.~B.~Dent, S.~Dutta and E.~N.~Saridakis,
  ``Cosmological perturbations in f(T) gravity,''
  arXiv:1008.1250 [astro-ph.CO].



\bibitem{Bengochea:2010sg}
  G.~R.~Bengochea,
  ``Observational information for f(T) theories and Dark Torsion,''
  arXiv:1008.3188 [astro-ph.CO].

\bibitem{Wu:2010av}
  P.~Wu and H.~W.~Yu,
  ``$f(T)$ models with phantom divide line crossing,''
  arXiv:1008.3669 [gr-qc].

\bibitem{Bamba:2010iw}
  K.~Bamba, C.~Q.~Geng and C.~C.~Lee,
  ``Comment on 'Einstein's Other Gravity and the Acceleration of the
  Universe'',''
  arXiv:1008.4036 [astro-ph.CO].

\bibitem{Karami:2010bu}
  K.~Karami and A.~Abdolmaleki,
  ``Original and entropy-corrected versions of the holographic and new
  agegraphic f(T)-gravity models,''
  arXiv:1009.2459 [gr-qc].

\bibitem{Karami:2010xy}
  K.~Karami and A.~Abdolmaleki,
  ``Reconstructing f(T)-gravity from the polytropic and different Chaplygin gas
  dark energy models,''
  arXiv:1009.3587 [physics.gen-ph].

\bibitem{Dent:2010bp}
  J.~B.~Dent, S.~Dutta and E.~N.~Saridakis,
  ``f(T) gravity mimicking dynamical dark energy. Background and perturbation
  analysis,''
  arXiv:1010.2215 [astro-ph.CO].


\bibitem{Ferraro:2006jd}
  R.~Ferraro and F.~Fiorini,
  ``Modified teleparallel gravity: inflation without inflaton,''
  Phys.\ Rev.\  D {\bf 75}, 084031 (2007)
  [arXiv:gr-qc/0610067];\\
  R.~Ferraro and F.~Fiorini,
  ``On Born-Infeld Gravity in Weitzenbock spacetime,''
  Phys.\ Rev.\  D {\bf 78}, 124019 (2008)
  [arXiv:0812.1981 [gr-qc]].


\bibitem{new GR}
  K.~Hayashi and T.~Shirafuji,
  ``New general relativity,''
  Phys.\ Rev.\  D {\bf 19}, 3524 (1979)
  [Addendum-ibid.\  D {\bf 24}, 3312 (1982)].



\bibitem{Wang:1998gt}
  L.~M.~Wang and P.~J.~Steinhardt,
  ``Cluster Abundance Constraints on Quintessence Models,''
  Astrophys.\ J.\  {\bf 508}, 483 (1998)
  [arXiv:astro-ph/9804015].


\bibitem{Starobinsky:1998fr}
  A.~A.~Starobinsky,
  ``How to determine an effective potential for a variable cosmological
  term,''
  JETP Lett.\  {\bf 68}, 757 (1998)
  [Pisma Zh.\ Eksp.\ Teor.\ Fiz.\  {\bf 68}, 721 (1998)]
  [arXiv:astro-ph/9810431].


\bibitem{Linder:2003dr}
  E.~V.~Linder and A.~Jenkins,
  ``Cosmic Structure and Dark Energy,''
  Mon.\ Not.\ Roy.\ Astron.\ Soc.\  {\bf 346}, 573 (2003)
  [arXiv:astro-ph/0305286].


\bibitem{Nojiri:2006ri}
  S.~Nojiri and S.~D.~Odintsov,
  ``Introduction to modified gravity and gravitational alternative for dark
  energy,''
  eConf {\bf C0602061}, 06 (2006)
  [Int.\ J.\ Geom.\ Meth.\ Mod.\ Phys.\  {\bf 4}, 115 (2007)]
  [arXiv:hep-th/0601213].



\bibitem{fR growth}
  S.~Tsujikawa,
  ``Matter density perturbations and effective gravitational constant in
  modified gravity models of dark energy,''
  Phys.\ Rev.\  D {\bf 76}, 023514 (2007)
  [arXiv:0705.1032 [astro-ph]].




\bibitem{LLT}
  B.~Li, T.~P.~Sotiriou and J.~D.~Barrow,
  ``f(T) gravity and local Lorentz invariance,''
  arXiv:1010.1041 [gr-qc]; \\
  E.~E.~Flanagan and E.~Rosenthal,
  ``Can Gravity Probe B usefully constrain torsion gravity theories?,''
  Phys.\ Rev.\  D {\bf 75}, 124016 (2007)
  [arXiv:0704.1447 [gr-qc]].

\bibitem{boundary}
  H.~I.~Arcos and J.~G.~Pereira,
  ``Torsion Gravity: a Reappraisal,''
  Int.\ J.\ Mod.\ Phys.\  D {\bf 13}, 2193 (2004)
  [arXiv:gr-qc/0501017].



\bibitem{mukhanov}
  V.~F.~Mukhanov, H.~A.~Feldman and R.~H.~Brandenberger,
  ``Theory of cosmological perturbations. Part 1. Classical perturbations. Part
  2. Quantum theory of perturbations. Part 3. Extensions,''
  Phys.\ Rept.\  {\bf 215}, 203 (1992).

\bibitem{hamilton}
A.~J.~S.~Hamilton, ``General Relativity, Black Holes, and
Cosmology.''





\end{thebibliography}
\end{document}